\documentclass[lettersize,letterpaper]{IEEEtran}
\usepackage{amsmath,amsfonts}
\usepackage{algorithmic}
\usepackage{algorithm}
\usepackage{array}
\usepackage{textcomp}
\usepackage{stfloats}
\usepackage{url}
\usepackage{verbatim}
\usepackage{graphicx}
\usepackage{cite}
\usepackage{bm}
\usepackage{subcaption}
\usepackage{booktabs} 
\usepackage{multirow}
\usepackage{amsthm} 
\usepackage{hyperref} 
\usepackage{cleveref} 
\usepackage{makecell}

\begin{document}
	
	\title{Integrated Sensing and Communication for Segmented Waveguide-Enabled Pinching Antenna Systems}
	
	\author{Qian Gao,~\IEEEmembership{Graduate Student Member,~IEEE,} 
		Ruikang Zhong,~\IEEEmembership{Member,~IEEE,} \\
		Hyundong Shin,~\IEEEmembership{Fellow,~IEEE,}
		Yuanwei Liu,~\IEEEmembership{Fellow,~IEEE}

		\thanks{Qian Gao is with the School of Electronic Engineering and Computer Science, Queen Mary University of London, London, E1 4NS, U.K. (e-mail: q.gao@qmul.ac.uk).
		
		Ruikang Zhong is with the  School of Information and Communication Engineering, Xi’an Jiaotong University, Xi’an 701149, P.R. China. (e-mail: ruikang.zhong@xjtu.edu.cn)
		
		Hyundong Shin is with the Department of Electronics and Information
		Convergence Engineering, Kyung Hee University, 1732 Deogyeong-daero,
		Giheung-gu, Yongin-si, Gyeonggi-do 17104, Republic of Korea. (e-mail:
		hshin@khu.ac.kr)
		
		Yuanwei Liu  is with
		the Department of Electrical and Electronic Engineering, The University
		of Hong Kong, Hong Kong. (e-mail: yuanwei@hku.hk).
	
}
	}

	\maketitle

\begin{abstract}

In this paper, an integrated sensing and communication (ISAC) design for segmented waveguide-enabled pinching-antenna array (SWAN) systems is proposed to improve the performance of systems by leveraging the low in-waveguide propagation loss of segmented waveguides. The hybrid segment selection and multiplexing (HSSM) protocol is implemented to provide favorable performance with less hardware cost. To achieve this, a joint transmit beamforming optimization, segment selection, and pinching antenna positioning problem is formulated to maximize the sum communication rate with the constraints of sensing performance. To solve the maximization problem, we propose a segment hysteresis based reinforcement learning (SHRL) algorithm to learn segment selection and pinching antenna positions in different progress to explore better strategies. Simulation results demonstrate that 1) the proposed SWAN-ISAC scheme outperforms the other baseline schemes, and 2) the proposed HARL algorithm achieves better performance compared to conventional RL algorithms.

\end{abstract}


\section{Introduction}

The rapid development of sixth-generation (6G) wireless networks is expected to enable a wide range of intelligent services such as autonomous driving, extended reality, and digital twin systems. To support these applications, networks must not only provide ultra-reliable high-rate communication but also acquire fine-grained awareness of the surrounding environment. Integrated Sensing and Communication (ISAC) \cite{ISAC6G} has therefore emerged as a key paradigm that allows communication and radar sensing to share the same hardware platform, spectrum, and signaling resources. By transmitting dual-functional waveforms, base stations can simultaneously deliver data to mobile users and probe the environment for target detection, localization, and tracking. This integration not only improves spectrum efficiency but also promotes tighter cooperation between communication and sensing functions, which is crucial for vehicular networks, unmanned aerial systems, and industrial Internet-of-Things (IoT) scenarios \cite{uav}.

A fundamental challenge for ISAC lies in the antenna hardware. Traditional phased-array or reconfigurable antennas \cite{fuild,move} can steer beams electronically, but they often suffer from high cost, complex feeding networks, and limited flexibility when adapting to dynamically changing user and target distributions. Recently, the concept of pinching antenna \cite{PA} has been introduced as a promising candidate for ISAC systems. A pinching antenna can be realized by directly “pinching” the waveguide or transmission line to form a radiation point. This design provides high spatial flexibility, as radiation points can be created at arbitrary positions along the waveguide. Compared with other types of flexible antennas, such as movable antennas\cite{move} or fluid antennas\cite{fuild}, the pinching antenna exhibits lower hardware complexity, more efficient energy transfer, and the capability of forming multiple radiation points without complicated mechanical movement. These features make pinching antennas become an attractive enabler for ISAC, as they can improve the flexibility of joint optimization of beamforming and sensing coverage while reducing system cost.

However, existing pinching-antenna designs \cite{PA1Ddl,PA1Dul,PA2D} are implemented on single or multiple tens-of-meters-long waveguides, which degrade system performance when covering a large region and lead to costly maintainability. To further enhance the flexibility of pinching antennas, recent studies have proposed the use of segmented waveguide-enabled pinching-antenna array (SWAN) systems \cite{seg}. In this architecture, the waveguide is physically divided into multiple segments, each of which can be selectively activated to serve as a controllable radiation aperture. The segmented design introduces a new degree of freedom, allowing more refined spatial resource allocation and reducing in-waveguide propagation loss.  

Motivated by these benefits, this paper proposes the ISAC design for SWAN. Specifically, the hybrid segment selection and multiplexing (HSSM) protocol is implemented, where each segment is connected to one dedicated radio frequency (RF) chain and segments are selected to realize multiplexing with less hardware cost. Then, the ISAC performance can be maximized by optimizing the transmit beamforming, segment selection, and pinching antenna positions. A segment hysteresis based reinforcement learning (SHRL) algorithm is proposed to tackle the non-convex optimization problem.


\section{System Model and Problem Formulation} \label{section:2}

\begin{figure*}[t!]
	\centering
	\captionsetup{justification=centering}
	\includegraphics[width=0.7\textwidth]{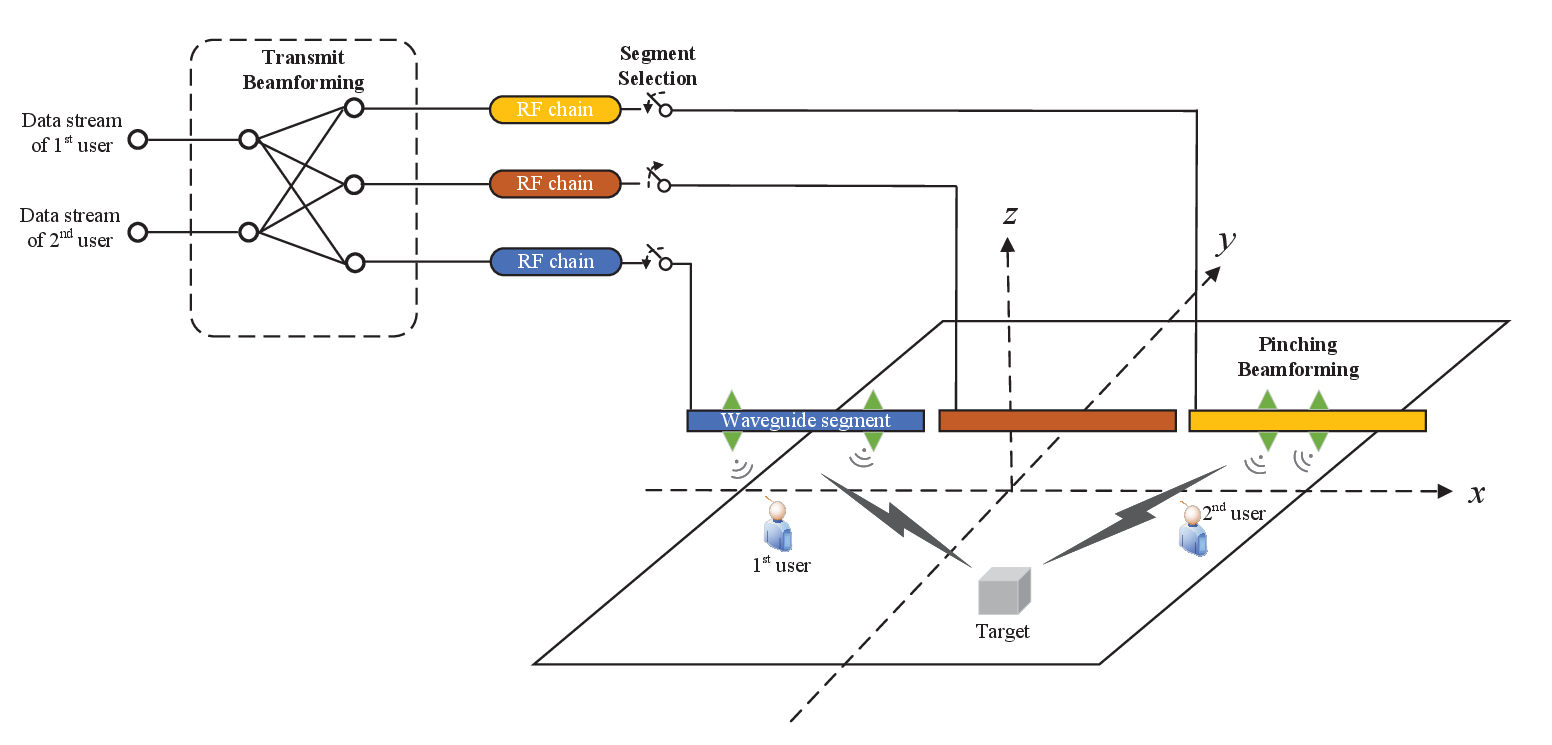}
	\caption{Illustration of SWAN-ISAC, $M =3, N = 4, K_c = 2, K_s = 1$.}
	\label{Fig.1}
\end{figure*}

In this paper, we investigate the downlink SWAN-ISAC system, as shown in Fig. \ref{Fig.1}. We assume that base station~(BS) is equipped with $M$ segmented waveguides, each with length $L$, to jointly serve $K_c$ single-antenna communication users and detect $K_s$ sensing targets. Each waveguide contains $N$ pinching antennas~(PAs), resulting in $M \times N$ total antennas. The positions of the $n$th PA on the $m$th waveguide can be denoted as $\bm{\psi}^{m}_{n} = (x^{m}_n, y^{m}_n, d)$, where $d$ is the fixed waveguide height above the $x$-$y$ plane. The feed point of each waveguide is assumed to be the front-left end of each waveguide. Communication users and sensing targets are located in the $x$-$y$ plane, with coordinates $\bm{\psi}_{k_c} = (x_{k_c}, y_{k_c}, 0)$ and $\bm{\psi}_{k_s} = (x_{k_s}, y_{k_s}, 0)$, respectively.

\subsection{Signal Model}

Compared to the time-division multiple access (TDMA) scheme used in single waveguide and multiple systems, the segmented waveguides can be connected to dedicated RL chains to enable maximal ratio transmission (MRT) in the downlink. The transmitted signal for the $m$th waveguide is denoted as
\begin{equation}
	s_m = \sum_{k_c=1}^{K_c} \mathbf{W}_{k_c, m} \cdot z_{k_c},
\end{equation}
where $z_{k_c}$ is the transmitted signal symbol for the $k_c$ th user, $\mathbf{W} \in \mathcal{C}^{K_c\times M}$ is the transmit beamforming.

Then, this signal is fed into the corresponding waveguide for pinching beamforming. For each waveguide, the signals transmitted by PAs are phase-shifted versions of each other. Considering the in-waveguide loss, the radiation signal can be expressed as

\begin{equation}
	\mathbf{s}_m = s_m \left[
	\begin{array}{c}
		\sqrt{\tfrac{\text{P}_m}{N}}\,10^{-\tfrac{\kappa}{20}\|\bm{\psi}^{m}_{0}-\bm{\psi}^{m}_{1}\|}e^{-j\theta^{m}_{1}} \\
		\vdots \\
		\sqrt{\tfrac{\text{P}_m}{N}}\,10^{-\tfrac{\kappa}{20}\|\bm{\psi}^{m}_{0}-\bm{\psi}^{m}_{N}\|}e^{-j\theta^{m}_{N}}
	\end{array}
	\right]^{\text{T}}
\end{equation}
where $\text{P}_m$ is the feed power on $m$th waveguide, $\theta^{m}_{n}$ is the signal phase shift at $n$-th PA and $\theta^{m}_{n} = 2\pi \frac{|\bm{\psi}^{m}_{0}-\bm{\psi}^{m}_{n}|}{\lambda_0}$. Thereof, $\bm{\psi}^{m}_{0}$ is the feed point of $m$th waveguide, $\lambda_0 = \frac{\lambda}{n_{\text{neff}}}$ is the waveguide wavelength in a dielectric waveguide, $n_{\text{neff}}$ is the effective refractive index and $\lambda$ represents wavelength. It is assumed that transmit power is allocated to $N$ pinching antennas evenly and this results in the scaling factor of $\sqrt{\tfrac{1}{N}}$.

Since the employment of PAs provides LoS channel conditions for ISAC, the near-field spherical wave channels for communication user $k_c$ and sensing target $k_s$ can be expressed as 
\begin{align}
	\mathbf{h}_{k_c,m} = \left[\frac{\alpha e^{-j \frac{2\pi}{\lambda}||\bm{\psi}_{k_c} -\bm{\psi}^{m}_{1}||}}{||\bm{\psi}_{k_c}  -\bm{\psi}^{m}_{1}||}, \cdots,
	\frac{\alpha e^{-j \frac{2\pi}{\lambda}||\bm{\psi}_{k_c}  -\bm{\psi}^{m}_{N, t}||}}{||\bm{\psi}_{k_c} -\bm{\psi}^{m}_{N}||}\right]^\text{T},
\end{align}
and
\begin{align}
	\mathbf{h}_{k_s,m} = \left[\frac{\alpha e^{-j \frac{2\pi}{\lambda}||\bm{\psi}_{k_s} -\bm{\psi}^{m}_{1}||}}{||\bm{\psi}_{k_s}  -\bm{\psi}^{m}_{1}||}, \cdots,
\frac{\alpha e^{-j \frac{2\pi}{\lambda}||\bm{\psi}_{k_s}  -\bm{\psi}^{m}_{N, t}||}}{||\bm{\psi}_{k_s} -\bm{\psi}^{m}_{N}||}\right]^\text{T},
\end{align}
where $\alpha = \frac{c}{4\pi f_c}$ is a constant that depends on the speed of light $c$ and the carrier frequency $f_c$, $\psi_{k,t}$ and $\psi_{l,t}$ denote the location of user $k$ and target $l$ in time slot $t$, respectively.

\subsection{Communication and Sensing Metrics}

The received signal at communication user $k_c$ is $y_{k_c} = \sum_{m=1}^{M}\mathbf{h}_{k_c,m}^\text{T}\mathbf{s}_m + \sum_{m=1}^{M}\sum_{k_c'=1, k_c' \neq k_c}^{K_c}\mathbf{h}_{k_c,m}^\text{T}\mathbf{s}_m' + n$, where $n$ represents the additive Gaussian noise.

Then, the data rate of user $k$ can be expressed as
\begin{align}
	R_{k_c} &= B \log_2 \left(1+ \frac{\sum_{m=1}^{M}|\mathbf{h}_{k_c,m}^\text{T}\mathbf{s}_m|^2}{\sum_{m=1}^{M}\sum_{k_c'=1, k_c' \neq k_c}^{K_c}|\mathbf{h}_{k_c,m}^\text{T}\mathbf{s}_m'|^2 + \sigma^2}\right),
\end{align}
where $\sigma^2$ is the noise power and $B$ represents bandwidth.  

The communication signals are further exploited for targets' sensing. The illumination power is used as the sensing performance metric, which can be expressed as

\begin{align} 
	\Gamma_{k_{s}} &=\mathbb{E}\left [\sum_{m=1}^{M}|\mathbf{h}_{k_s,m}^\text{T}\mathbf{s}_m|^2\right] .
\end{align}

To fulfill the requirements of sensing quality, the illumination towards different sensing targets needs to consider the minimum threshold $\tilde{\Gamma}$.

\subsection{Problem Formulation}

Based on the above system model, we formulate the joint optimization of the pinching beamforming $\{\bm{\psi}^m_{n}\}$, the segment selection $\{\phi_m\}$, and the transmit beamforming $\{\mathbf{W}_{k_c, m}\}$ to maximize the total communication data rate across all users while satisfying sensing threshold.
The optimization problem is:
\begin{subequations} \label{eq:opt_problem}
	\begin{align}
		\max_{\{\bm{\psi}^m_{n}, \phi_m, \mathbf{W}_{k_c, m}\}} &\quad  \sum_{k_c=1}^{K_c} R_{k_c} \notag \\
		\text{s.t.}\quad 
		&\quad \Gamma_{k_s} \geq \tilde{\Gamma}, \quad \forall k_s \in \{1,...,K_s\},  \\
		&\quad   |\mathbf{W}|^2 = \sum_{m=1}^{M}\text{P}_m \leq \text{P},\forall m,\\
		&\quad   \phi_m \in [0,1],\forall m,\\
		&\quad \|\bm{\psi}^{m}_{n} - \bm{\psi}^{m}_{n'}\| \geq \delta, \quad \forall n \neq n', \forall m,n. 
	\end{align}
\end{subequations}
Constraint (7a) ensures the minimum SNR requirement for target sensing, constraints (7b) and (7c) enforce the total
transmit power and the range of segment selection, respectively. Constraint (7d) guarantees the minimum spacing between any two pinching antennas. This optimization problem is nonlinear and non-convex due to the presence of fractional illumination power expressions, absolute values, and exponential phase terms.

\section{Proposed Segment Hysteresis based Reinforcement Learning Solution}\label{section:3}

In this section, we introduce the Segment-Hysteresis Reinforcement Learning (SHRL) 
framework, which is specifically designed for ISAC systems employing segmented 
waveguides.  
Different from conventional RL approaches that update segment selections independently at 
each step, SHRL introduces a hysteresis-based segment-update mechanism that 
stabilizes the agent's decisions by preventing abrupt changes in the antenna–segment 
mapping. This mitigates the non-stationarity caused by rapid structural switching 
and significantly improves learning stability.

The core idea of SHRL is to integrate a probabilistic hysteresis gate into the 
segment-selection process so that new segment decisions are only accepted when they 
provide sufficiently large improvement or when a certain hysteresis threshold is 
exceeded. Otherwise, previous selections are retained. This mechanism enforces 
temporal consistency and enables smooth policy exploration in the SWAN-ISAC 
optimization task.

\subsection{MDP Formulation with Segment Hysteresis}

The SHRL framework operates within a single unified environment 
$\mathcal{E}$ that comprises all antenna elements and segmented waveguide modules.  
The MDP is defined as follows.

\textbf{State Space $\mathcal{S}$:}
At each time step $t$, the system state is represented by
\[
\bm{s}_t = [\mathbf{h}_{k_c,t}, \mathbf{h}_{k_s,t},
\bm{\psi}_{t-1}, \bm{\phi}_{t-1}],
\]
including the communication and sensing CSI, the previous antenna positions 
$\bm{\psi}_{t-1}$, and the 
previous segment assignments $\bm{\phi}_{t-1}$, which are essential for the 
hysteresis mechanism.

\textbf{Action Space $\mathcal{A}$:}
The agent outputs continuous-valued updates for
\[
\bm{a}_t = [\bm{\psi}_{t}, \bm{\phi}_{t}, \mathbf{W}_{t}],
\]
where $\bm{\phi}_{t}$ represents the raw segment-selection logits.  
Before executing $\bm{\phi}_{t}$, a hysteresis filter is applied to determine 
whether segment reassignment should occur.

\textbf{Reward Function $\mathcal{R}$:}
The reward encourages high communication throughput while satisfying sensing-SNR 
constraints:
\[
r_t = \sum_{k_c=1}^{K_c} R_{k_c,t}
- \sum_{k_s=1}^{K_s}\mathbb{I}_{\{\Gamma_{k_s,t}<\tilde{\Gamma}\}}.
\]

\subsection{Segment Hysteresis Mechanism}

Given the raw segment logits $\bm{\phi}_{t}$ from the actor, SHRL determines the 
final segment assignment $\tilde{\bm{\phi}}_{t}$ according to:
\[
\tilde{\phi}_{m,t} =
\begin{cases}
	\phi_{m,t}, & \text{with probability } p_{\text{update}}, \\
	\tilde{\phi}_{m,t-1}, & \text{otherwise},
\end{cases}
\]
where $p_{\text{update}} \in (0,1)$ is the hysteresis probability controlling the 
update frequency. A small $p_{\text{update}}$ preserves stability, while a larger 
value allows more aggressive exploration.  
This mechanism prevents frequent remapping of antennas across segments, thereby 
reducing structural variance and improving the convergence behavior of the policy.

\subsection{Training Procedure}

We employ the Advantage Actor–Critic (A2C) algorithm~\cite{A2C} to optimize the 
SHRL policy. The actor outputs both beamforming parameters and raw segment logits, 
while the hysteresis gate refines the segment-selection action before it is applied 
to the environment. The critic evaluates state values and guides policy updates.

After training convergence, the learned policy 
$\pi_{\theta^*}$ is used for real-time control of antenna positions, beamforming weights, and hysteresis-smoothed segment selection.

\begin{algorithm}[ht]
	\caption{SHRL: Reinforcement Learning with Segment Hysteresis}
	\label{alg:shrl}
	\begin{algorithmic}[1]
		\STATE Initialize policy $\pi_\theta$, critic $V_\psi$ and hysteresis memory $\tilde{\bm{\phi}}_0$.
		\FOR{each episode}
		\STATE Reset environment and observe initial state $s_0$.
		\FOR{each step $t$}
		\STATE Sample raw action $a_t = \pi_\theta(s_t)$, extract raw segment logits $\bm{\phi}_t$.
		\STATE Apply hysteresis gate:
		\STATE $\tilde{\bm{\phi}}_t =
		\left\{
		\begin{aligned}
			&\bm{\phi}_t, && \text{with prob. } p_{\text{update}},\\
			&\tilde{\bm{\phi}}_{t-1}, && \text{otherwise}.
		\end{aligned}
		\right.$
		\STATE Execute action $(\bm{\psi}_t, \tilde{\bm{\phi}}_t, \mathbf{W}_t)$ in the environment.
		\STATE Observe $r_t$ and $s_{t+1}$.
		\STATE Update critic and actor using A2C:
		\STATE $\mathcal{L}_V = (V_\psi(s_t) - \hat{R}_t)^2,\quad
		\mathcal{L}_\pi = \log\pi_\theta(a_t|s_t)\hat{A}_t.$
		\STATE Store updated $\tilde{\bm{\phi}}_t$.
		\ENDFOR
		\ENDFOR
		\STATE Output the final policy parameters $\theta^*$.
	\end{algorithmic}
\end{algorithm}

The SHRL approach provides an effective balance between exploration efficiency and 
structural stability, making it particularly suitable for large-scale modular ISAC 
architectures. By smoothing segment transitions through hysteresis, SHRL achieves 
more stable convergence compared with conventional RL baselines, while 
maintaining high communication–sensing performance.

\section{Simulation Results}\label{section:4}

In this section, we conduct experimental simulations for the proposed SWAN-ISAC deployment and SHRL algorithm. In the simulation, we consider a multi-user ISAC system with a configurable pinching antenna deployment geometry, where $U = 6$ users and $T = 1$ target are randomly distributed within a $50\,\mathrm{m} \times 60\,\mathrm{m}$ area at ground level. The antenna array of each segment consists of $N = 10$ elements. The carrier frequency is set to $f = 28\,\mathrm{GHz}$, and each antenna element is allowed to adjust its position under a minimum inter-element spacing $\delta =\lambda/2$. Communication channels are modeled as LoS paths with effective refractive index $n_{\mathrm{eff}} = 1.4$. The total power constraint is $P_{\max} = 100\,\mathrm{W}$ and the height of waveguides is 5$\mathrm{m}$. The communication data rate is computed based on coherent beamforming with additive white Gaussian noise power of $-90\,\mathrm{dBm}$, and the sensing performance is evaluated via the maximum illumination power achieved for the target. The sensing threshold $\tilde{\Gamma}$ is set to $-20\,\mathrm{dBm}$.

To evaluate the performance of the proposed SHRL algorithm, we adapt four benchmarks as follows:
\begin{itemize}
	\item SPRL: Segment periodical selection is a variant of the proposed SHRL algorithm, where segment selection are made in a fixed period. In this paper, we got period 5 with grid search.
	\item A2C: This is a synchronous policy-gradient algorithm where a single actor selects actions and a critic estimates the value function. It updates the policy using the advantage signal.
	\item PPO: PPO improves policy-gradient updates by constraining the step size using a clipped surrogate objective. This prevents overly large policy updates and provides better stability and robustness than vanilla actor–critic or A2C, especially in high-dimensional or multi-agent settings.. 
	\item Random scheme: This scheme requires the agent to execute random actions within predefined regions.
\end{itemize}

\begin{figure}[t!]
	\centering
	\captionsetup{justification=centering}
	\includegraphics[width=0.5\textwidth]{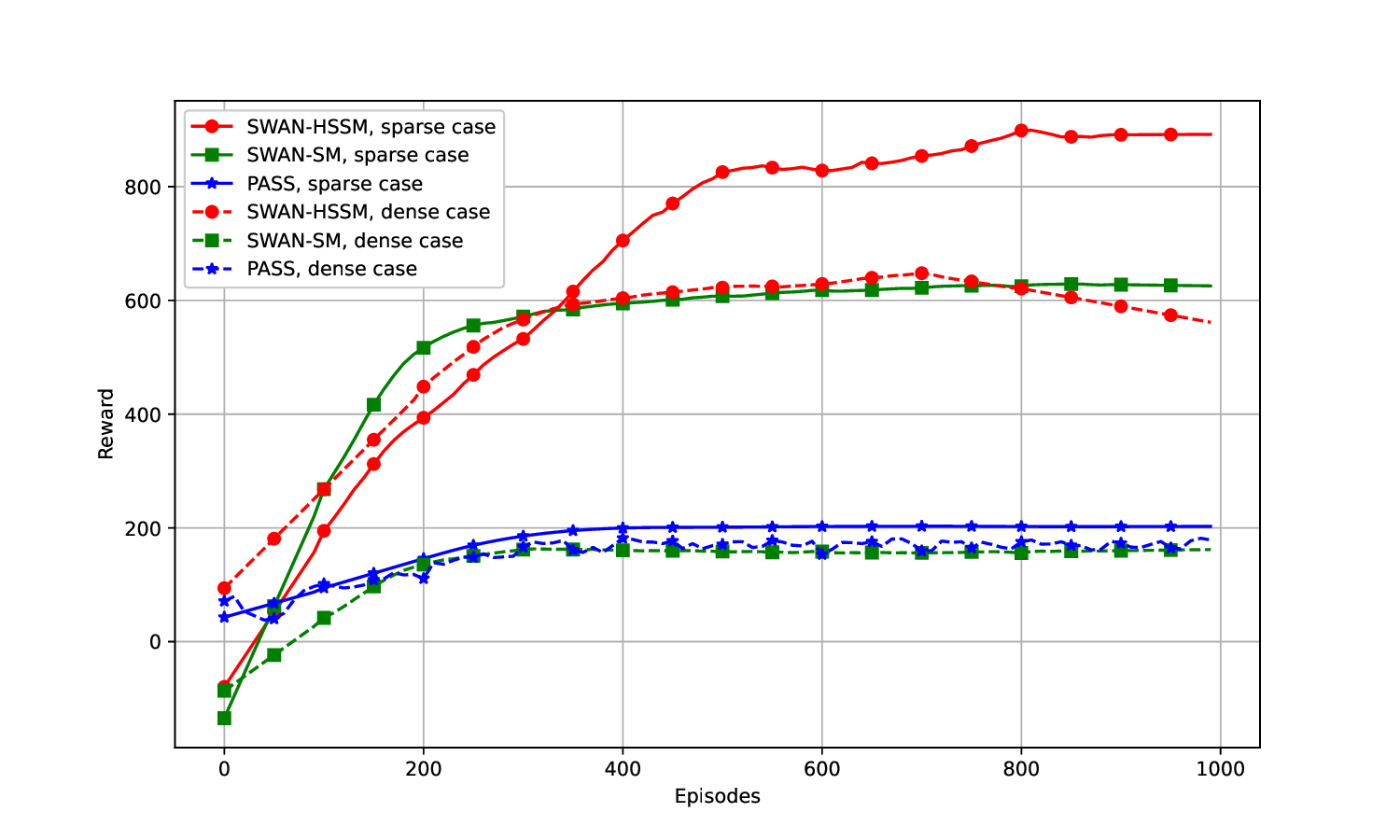}
	\caption{ISAC performance comparison between SWAN and PASS under two different cases.}
	\label{Fig.2}
\end{figure}

Fig.~\ref{Fig.2} compares the ISAC performance of the SWAN (HSSM and SM protocols) against the conventional PASS scheme under two representative cases. In the sparse case, where users and targets are spatially separated, both SWAN-HSSM and SWAN-SM achieve significantly higher rewards than PASS. SWAN-HSSM exhibits the highest final reward, owing to its ability to selectively activate effective segments and reduce unnecessary power spreading. SWAN-SM also performs well but remains slightly inferior due to the lack of hybrid-selective control. In contrast, PASS yields limited performance because its power allocation mechanism does not exploit spatial diversity.

In the dense case, where users and targets are located close to each other, the reward of all schemes decreases. This degradation is attributed to stronger multi-user interference and reduced beamforming separability. However, SWAN-HSSM still maintains a notable advantage over SWAN-SM and PASS, as selective segment activation helps mitigate propagation loss and suppress interference. PASS continues to show the lowest performance across both cases due to its fixed, non-adaptive structure.

\begin{figure}[t!]
	\centering
	\captionsetup{justification=centering}
	\includegraphics[width=0.5\textwidth]{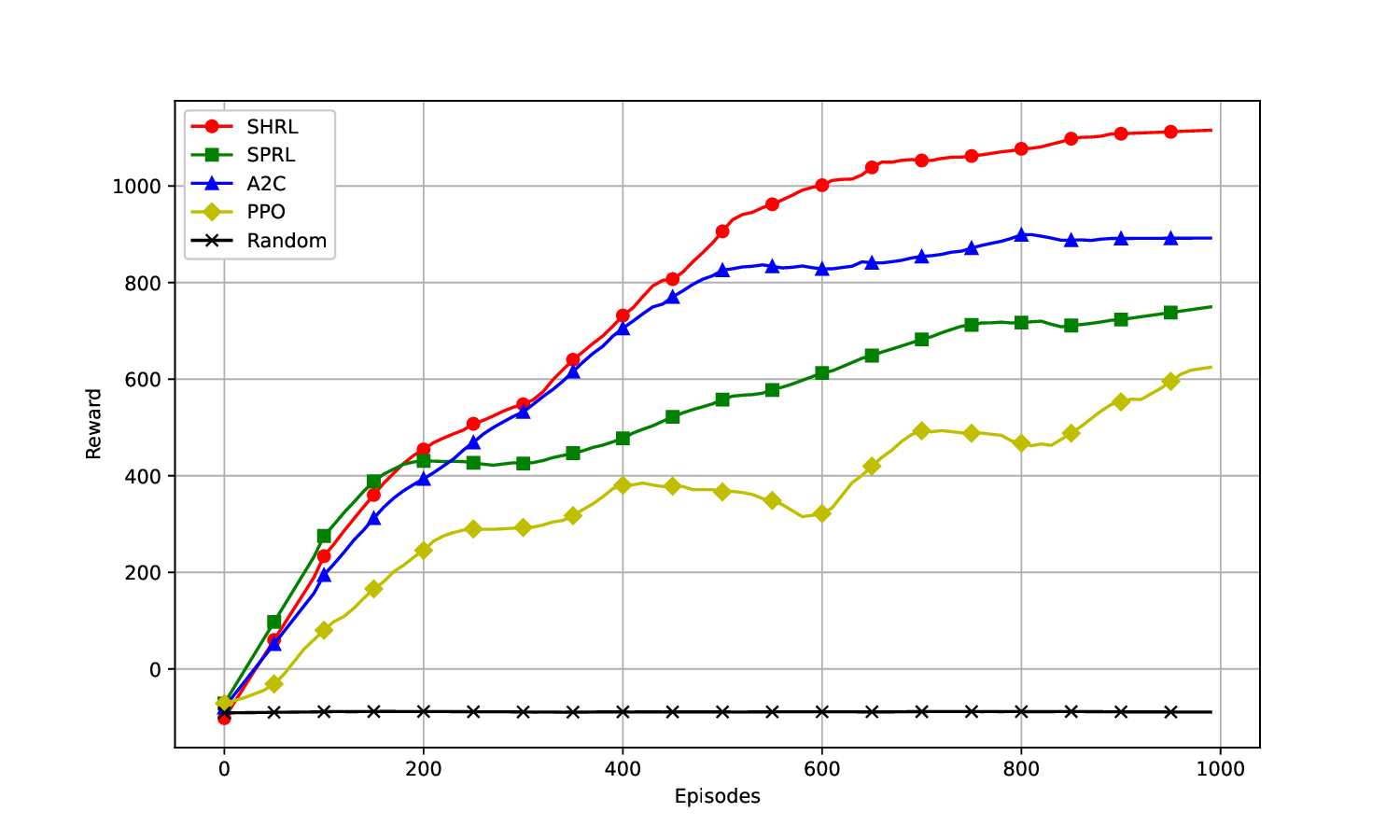}
	\caption{Performance comparison of SWAN-ISAC under different algorithms.}
	\label{Fig.3}
\end{figure}

Fig.~\ref{Fig.3} compares the learning performance of different reinforcement 
learning strategies for SWAN-ISAC. SHRL, which introduces probabilistic hysteresis for 
segment updating, achieves the highest reward and the most stable convergence by 
preventing unnecessary segment switching. A2C ranks second, benefiting from more 
consistent policy updates in this continuous control setting.
SPRL, which updates segments only at fixed intervals, converges more slowly and reaches 
a lower performance ceiling due to reduced exploration flexibility. PPO performs worse 
because its clipping mechanism over-constrains policy updates in this multi-parameter 
environment. As expected, the Random policy fails to learn any meaningful behavior.
Overall, SHRL provides the most effective balance between exploration stability and 
segment-control flexibility, leading to the best ISAC performance.

\begin{table*}[t]
	\centering
	\caption{ISAC performance comparison of different algorithms under varying
		waveguide configurations.}
	\label{tab:isac_waveguide}
	
	\begin{subtable}{\linewidth}
		\centering
		\caption{Fixed segment number $M=3$, varying total waveguide length $L_{\mathrm{wg}}$.}
		\label{tab:isac_fixedS}
		
		\begin{tabular}{c|ccccc}
			\hline
			$L_{\mathrm{wg}}$ & SHRL & SPRL & A2C & PPO & Random \\
			\hline
			40\,$\mathrm{m}$ &  $13.15/1.09e-5$ &
			$ 7.71/1.05e-5 $ &
			$9.93/1.12e-5$ &
			$6.37/1.18e-5$ &
			$3.05/9.19e-6$ \\
			60\,$\mathrm{m}$ &  $13.56/1.26e-5$ &
			$11.20/1.28e-5$ &
			$11.63/1.24e-5$ &
			$9.45/1.30e-5$ &
			$3.09/3.11e-6$ \\
			80\,$\mathrm{m}$ &  $17.07/1.34e-5$ &
			$11.32/1.42e-5$ &
			$13.85/1.37e-5$ &
			$10.86/1.45e-5$ &
			$3.55/2.77e-6$ \\
			100\,$\mathrm{m}$ &  $6.41/1.10e-5$ &
			$3.36/1.23e-5$ &
			$4.58/1.17e-5$ &
			$4.55/1.26e-5$ &
			$2.95/1.81e-6$ \\
			\hline
		\end{tabular}
	\end{subtable}

	\begin{subtable}{\linewidth}
		\centering
		\caption{Fixed total waveguide length $L_{\mathrm{wg}}=40 m$, varying segment number $M$.}
		\label{tab:isac_fixedL}
		
		\begin{tabular}{c|ccccc}
			\hline
			$M$ & SHRL & SPRL & A2C & PPO & Random \\
			\hline
			3 & $13.15/1.09e-5$ &
			$7.71/1.05e-5$ &
			$9.93/1.12e-5$ &
			$6.37/1.18e-5$ &
			$3.05/9.19e-6$ \\
			6 & $12.91/2.74e-5$ &
			$6.64/2.28e-5$ &
			$8.97/1.76e-5$ &
			$5.98/1.54e-5$ &
			$3.03/5.92e-6$ \\
			9 & $11.82/3.72e-5$ &
			$6.36/2.42e-5$ &
			$8.32/2.02e-5$ &
			$5.64/1.68e-5$ &
			$2.83/4.44e-6$ \\
			12 &$ 10.59/4.96e-5$ &
			$5.42/3.54e-5$ &
			$7.63/2.64e-5$ &
			$5.15/1.78e-5$ &
			$3.00/3.96e-6$ \\
			\hline
		\end{tabular}
	\end{subtable}
\end{table*}

Table~\ref{tab:isac_waveguide} summarizes the ISAC performance of different algorithms 
under varying waveguide configurations. Each entry in the table is presented in the form ``$R / \Gamma$'', 
where the first value denotes the total communication rate (in bps/Hz), 
and the second value denotes the achieved sensing performance (in dB). In Table~\ref{tab:isac_waveguide}(a), the 
segment number is fixed to $M=3$ while the total waveguide length $L_{\mathrm{wg}}$ 
varies. Across all configurations, SHRL consistently achieves the highest communication 
rate and maintains sensing SNR closest to the required threshold. As $L_{\mathrm{wg}}$ 
increases from 40\,m to 80\,m, all methods exhibit performance improvements due to the 
larger available spatial aperture. However, for very long waveguides 
(e.g., $L_{\mathrm{wg}}=100$\,m), the propagation loss becomes innegligible, resulting in 
performance degradation across all algorithms. SHRL remains the most robust in this 
high-loss regime, whereas SPRL and PPO suffer more substantial degradation. Random 
selection performs poorly in all cases, confirming the necessity of intelligent segment 
control and beamforming.
Table~\ref{tab:isac_waveguide}(b) fixes the total waveguide length at 
$L_{\mathrm{wg}}=40$\,m and varies the segment number $M$. The sensing SNR improves with larger $M$ because shorter 
per-segment waveguides experience lower in-waveguide attenuation, leading to 
stronger coherent reflection at the target. Conversely, the communication rate 
decreases since more segments reduce the number of antennas per segment, 
degrading the effective array gain and weakening the communication beamforming. 
Among all methods, SHRL consistently achieves the best balance between sensing 
and communication performance. For the random scheme, its sensing illumination power cannot meet the minimum threshold, which guarantees the sensing metric to be considered during the training of other algorithms.

\section{Conclusion}\label{section:5}
In this paper, we proposed a SWAN-enhanced ISAC framework that exploits the spatial diversity of segmented waveguides to enable flexible beamforming and efficient separation between communication and sensing. By formulating the joint optimization of antenna positioning, segment assignment, and transmit beamforming as a Markov decision process, we developed a SHRL algorithm capable of stabilizing segment transitions while maintaining exploration efficiency.
Compared with existing baselines such as SPRL, A2C, and PPO, the proposed SHRL method achieves superior learning stability and higher ISAC performance, owing to its effective handling of segment update dynamics and task-specific structural priors. Simulation results demonstrate that the SWAN-ISAC architecture maintains the sensing SNR near its operational threshold while significantly boosting communication throughput. These findings highlight the advantages of segmented-waveguide antenna systems and validate their potential for high-performance, power-limited ISAC deployments.
\bibliographystyle{IEEEtran}
\bibliography{ref}

\end{document}